# Long-range Self-assembly via the Mutual Lorentz Force of Plasmon Radiation

Haojie Ji[1], Jacob Trevino[2,3,4], Raymond Tu[4,5], Ellen Knapp[5], James McQuade[1], Vitaliy Yurkiv[6], Farzad Mashayek[6], and Luat T. Vuong[1,2,4,*]

1 Queens College of the City University of New York, Department of Physics, Flushing, NY 11367

2 The Graduate Center of the City University of New York, Department of Physics, New York, NY 10016

3 The Graduate Center of the City University of New York, Department of Chemistry, New York, NY 10016

4 Advanced Science Research Center of the Graduate Center at the City University of New York, New York, NY 10031

5 City College of New York, Department of Chemical Engineering, New York, NY 10031

6 University of Illinois at Chicago, Department of Mechanical and Industrial Engineering, Chicago, IL 60607

*Correspondence to: Luat.Vuong@qc.cuny.edu

**Abstract:** Long-range interactions often proceed as a sequence of hopping through intermediate, statistically-favored events. Here, we identify a widely-overlooked mechanism for the mechanical motion of particles that arises from the Lorentz force and plasmon radiation. Even if the radiation is weak, the nonconservative Lorentz force produces stable locations perpendicular to the plasmon oscillation; over time, distinct patterns emerge. Experimentally, linearly-polarized light leads to the formation of 80-nm Au nanoparticles, perpendicularly-aligned, with lengths that are orders of magnitude greater than their plasmon near-field interaction. There is a critical intensity threshold and optimal concentration for observing self-assembly.

**Main Text**

**Introduction.**

Scientists are modeling the underlying physical mechanisms that couple plasmons with each other and their environment [1] [2] [3] [4]. The collective presence of plasmons also carries mechanical consequences [5] [6] [7] [8] [9] [10] [11] that are complex and tied to the numerous channels by which plasmons rapidly decay: plasmon decay produces heat and shifts electrochemical potentials, while any corresponding motion, in turn, may alter the plasmon excitation. Currently, the most widespread methods for self-assembling, patterning, and optically trapping metal nanostructures avoid the direct excitation of plasmons [12], however plasmons are efficiently leveraged in nanostructuring processes [13] [14] [15] [16] [17]. Predictive control of these dynamics —i.e., the knowledge to model, *a priori,* the mechanical behavior of materials in the presence of plasmons—would aid the large-area nanomanufacturing of hybrid materials and advance other photonic technologies.

Although most research in the field of plasmonics focuses on near-field interactions, we revisit the contribution of the radiation from plasmons. The corresponding Lorentz forces are not only significant, but lead to the distinct, polarization-dependent, long-range pattern formation of conducting nanoparticles (NPs). With mutual Lorentz interactions—the force arising between radiated fields and other plasmons—each NP experiences torque, attraction or repulsion. NPs minimize energy by settling at stable locations aligned with other NPs, perpendicular to the linear polarization of light.

Our results show that the mutual Lorentz interactions lead to predictable patterns and longer-range interactions orders of magnitude greater than previously expected. The laser-induced self-assembly of gold NPs on indium-tin-oxide-coated (ITO) substrates show chains of NPs as long as 200 microns, aligned in a direction perpendicular to the polarization of the illuminating laser. Concentration-dependent regimes and intensity-dependent thresholds for self-assembly arise because of competition between mutual Lorentz forces and gradient or Brownian forces. These interactions between conducting NPs are not a conventional scattering or gradient force and are not generally under consideration [18] [19]; however, our results are consistent with prior studies [20] [21] [22] and explain recent experimental observations of the perpendicular alignment of NPs.

**Results.**

We consider the forces between two oscillating dipoles $p_1$ and $p_2$ positioned in the *x-y* plane where the dipoles are illuminated with a uniform plane wave traveling in the *z*-direction at the plasmon-resonant frequency. The dipoles carry uniform strength and equal frequency, and oscillate virtually in-phase with each other and the illuminating electric field [Fig. 1 (a)]. Prior work has scrutinized the interaction where the plasmon fields of the two NPs strongly overlap [23] [24] and where the distances between NPs are comparable to their diameter; the attractive interaction with NP dimers largely corresponds to a gradient force of the plasmon evanescent fields. We now consider longer-distance interactions that arise from the fields that radiate from each NP. Dipole $p_1$ produces an electromagnetic field that influences the other dipole $p_2$ and vice versa.

The mutual Lorentz force $F_{12} = F_{21}$ between two NPs with dipoles $p_1$ and $p_2$ is calculated,

$$F_{12} = (p_2 \cdot \nabla)E_{r1} + \frac{dp_2}{dt} \times B_{r1}, \tag{1}$$

where $E_{r1}$ and $B_{r1}$ are the radiated electric and magnetic fields from $p_1$ [as illustrated in Figs. 1(b-c)]. Although all three terms $p$, $E$, and $B$ oscillate as a function of time, the resultant force has a static term, a nonzero time-averaged contribution. It is well-known that the Lorentz force from a time-harmonic field may yield a static force, which is largely leveraged via light intensity gradients [20] [21] [22] [25]. However, under uniform light illumination where the gradient forces of the external fields vanish, the plasmon interactions described here dominate in their contribution of the NP self-assembly. The calculation of the static mutual Lorentz force is illustrated in Fig. 2(a) (See Supplemental Information for the derivation); the direction and strength of the force experienced by each dipole depends strongly on the fields at their relative positions. For 80-nm gold NPs, we estimate that the mutual Lorentz force, at distances comparable to a wavelength, is on the order of pN for 10 MW/cm$^2$ illumination intensities. This result indicates that most pulsed lasers may produce mutual Lorentz forces between plasmonic NPs that are on the order of the gradient forces conventionally leveraged in optical trapping experiments.

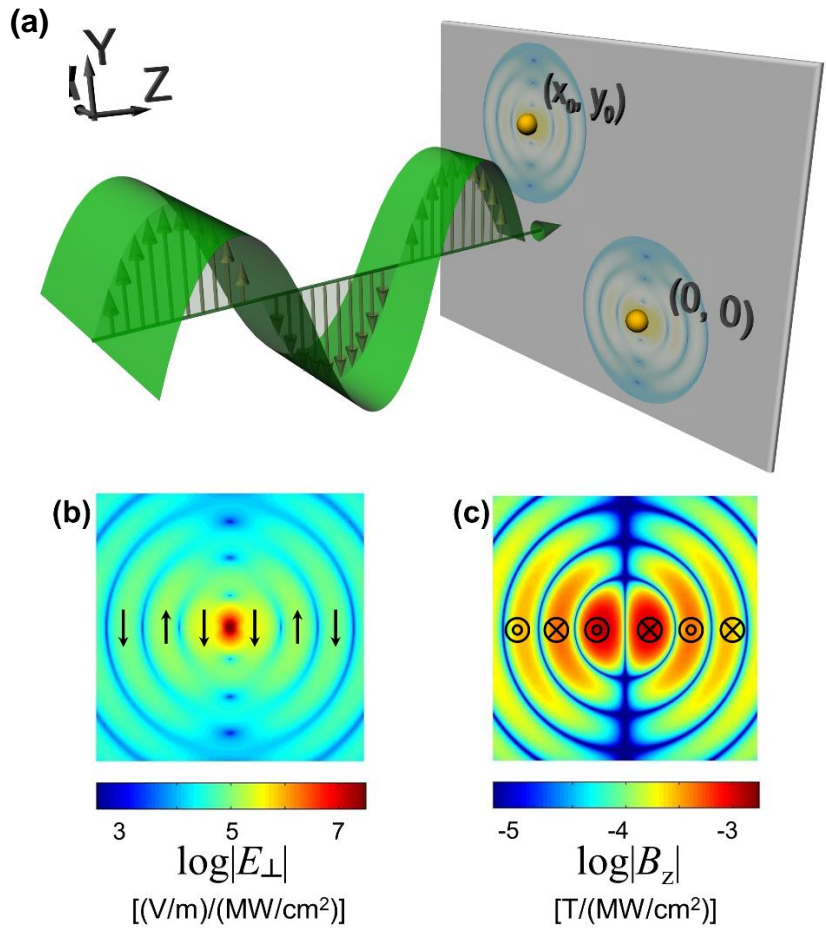

**Figure 1.** (a) General schematic describing Lorentz forces via dipole interactions: a plane wave with wavelength λ =

532 nm illuminates two in-phase oscillating dipoles whose displacement is ($x_0$, $y_0$). Each dipole radiates (b) electric $\boldsymbol{E}_r$ and (c) magnetic $\boldsymbol{B}_r$ fields in the *x-y* plane, shown for a 2-μm-length area. The color plots show the real part of the time-harmonic fields on a logarithmic scale.

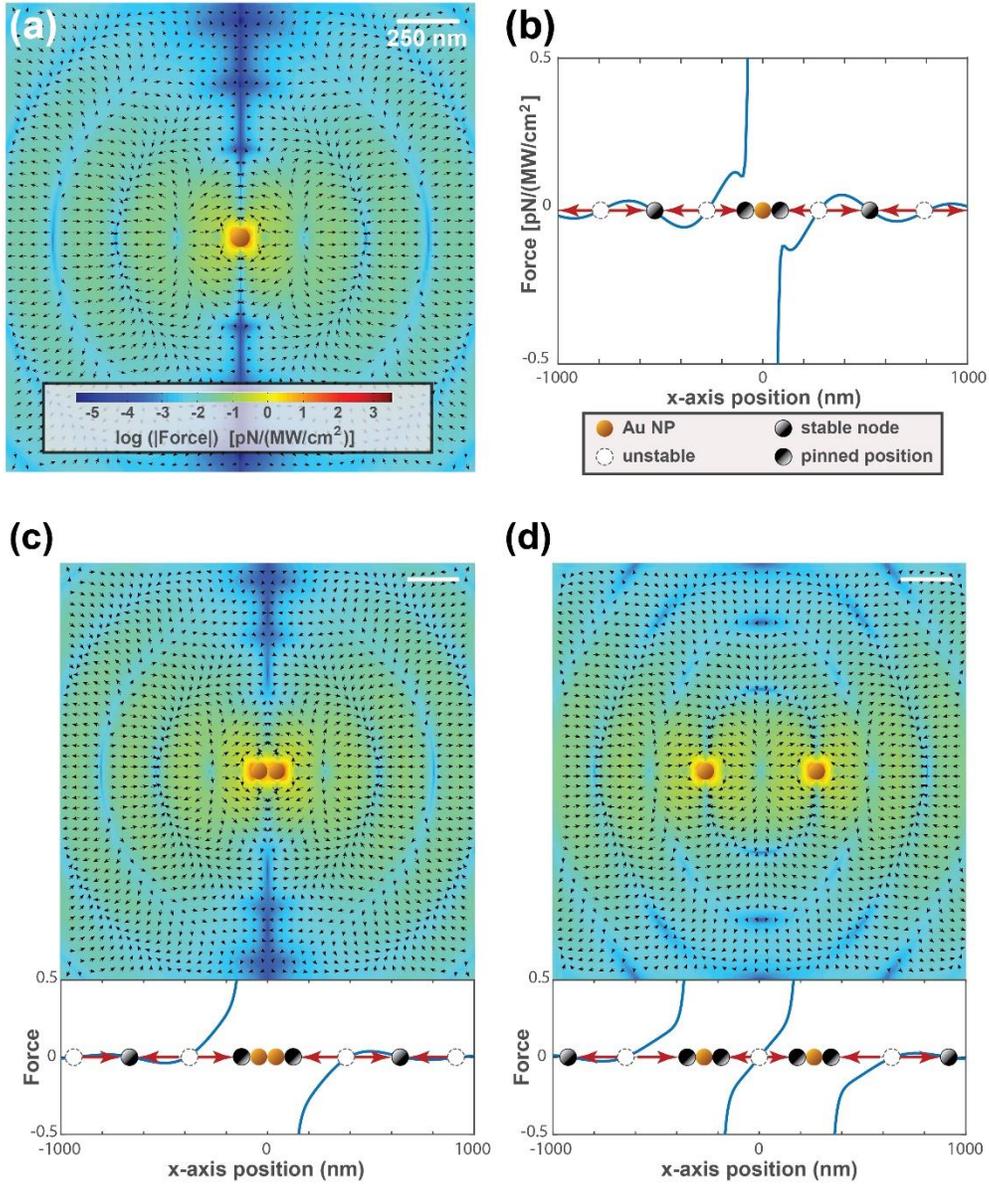

**Figure 2.** (a) Vector field showing the time-averaged mutual Lorentz force $F_{12}(x_0, y_0)$ experienced by an oscillating dipole positioned at ($x_0$, $y_0$) as a result of the dipole radiation from a nanoparticle (NP) at the origin. Calculations assume 532-nm-wavelength, vertically-polarized (*y*-direction) excitation and induced dipole from an 80-nm Au NP. (b) The corresponding stable, unstable, and stably-pinned positions on the *x*-axis as a result the gradient and mutual Lorentz forces. There is no vertical component; the force is pointed to right (left) if positive (negative). Similar vector-field plots of the mutual Lorentz force [above] and plot of the *x*-axis stable points [below] resulting from (c) 2 adjacent NPs and (d) 2 NPs separated by 532 nm on the *x*-axis.

The magnitude of the time-averaged Lorentz force decrease with $1/r^4$ in the near field and $1/r$ in the far field and is on the order of pN/(MW/cm$^2$) at distances comparable to a wavelength from the NP, which is significant for the plasmonic particle interaction. One particularly important, and unintuitive, characteristic of the vector field of this force: unlike the Coulomb interaction between electrostatic charges, the Lorentz force is not a conservative, "central" force and is not simply either attractive or repulsive; instead, Fig. 2 (a) shows that the time-averaged Lorentz force between oscillating dipoles may provide a torque between the dipoles. This feature is particularly unintuitive since the radiation momentum or Poynting of the radiated-field is directed radially outwards from each oscillating dipole. The mutual Lorentz force guides particles towards the $x$-axis in the far-field. In the near-field, the mutual Lorentz force is repulsive but weaker than the gradient force, which leads to the net attraction of NPs (see Supplemental Information for quantitative comparison).

The sum of the near-field gradient force and mutual Lorentz force produces two types of stable attractors for pairs of NPs in the $x$-$y$ plane [Fig. 2 (b)]. One of the stable attractors consists of pairs of adjacent NPs or dimers, pinned by a net attractive force and aligned perpendicular to the dipole oscillation or polarization (on the $x$-axis for $y$-polarized linear illumination). The other stationary attractor for the NPs consists of NPs spaced at integer-wavelength distances (again, on the $x$-axis for $y$-polarized linear illumination). Stable nodes appear at integer-wavelength separations and at these locations, the Lorentz force is directed inwards towards the nodes. The periodic forces along the $x$-axis can be viewed in a manner analogous to the Lorentz forces arising from electrical currents on parallel wires, which are either attractive or repulsive depending on whether the currents are aligned or opposed. Here, the direction of the force – either attractive or repulsive—is determined by the retarded phase of the radiated fields from adjacent NPs; the phase corresponds to the NP spacing. The resultant stable positions may have been observed in [26] for plasmonic NPs in an optical trap.

In a real NP system, Brownian forces are present. Over time, NPs move towards stable positions with lower energy over time. These new stable positions would likely be stable nodes that are closer or adjacent to other NPs, since the mutual Lorentz forces between NPs increases significantly with smaller separations, and gradient forces provide extra attraction for closer dimers [Fig. 2 (b)]. When multiple NPs are aligned close along the $x$-axis, the total energy is significantly lower. The calculation of the Lorentz force is subsequently extended to consider the superposition of two adjacent, or one-wavelength separated, NP dipoles [Fig. 2 (c) and (d)]. The vector flow of the forces and the presence of pinned positions and stable nodes on $x$-axis [Fig. 2 (c) and (d)] explain the subsequent 1-D alignments of NPs, which are reported in previous efforts [15] [16] [27], and observed in our own experiments.

In comparison to prior studies that have scrutinized the mechanical forces between plasmonic NPs, our results indicate that far-field effects contribute to the evolved system dynamics. Prior investigations have focused on how plasmon forces between pairs of nanoparticles influence their interactions in near-field distances [23] [24], and are associated with gradient forces [26] and thermal responses [28] [29] of the plasmon evanescent fields. In general, the near-field dimer interactions are attractive and may lead to the NP alignment either parallel or perpendicular to the illuminating laser polarization. In contrast, here we show that distinct signatures of the mutual Lorentz interactions are torque forces and the presence of stables nodes. We claim that the combination of these dynamics leads to large-scale perpendicular alignment of the NP chains, which depend on the incident electric-field polarization.

Figure 3(a) shows the schematic experimental setup and Fig. 3 (b-e) shows the SEM images of the NP drop-cast samples at different concentrations where the laser pulse peak intensities are 20MW/cm$^2$. Here, gold NPs settle on ITO-coated substrates under pulsed-laser illumination at the plasmonic resonance of 532 nm. Images are taken towards the center of the drops. The samples that are prepared without laser light and with circularly-polarized 532-nm laser light do not exhibit any patterns [Fig. 3 (b-c)]; by contrast, the settled Au NP samples prepared with linearly-polarized 532-nm laser light exhibit distinct string formations [Fig. 3 (d-e)]. With linearly-polarized 1060-nm (off resonant) laser light, no patterns are observed [not shown]. Alignment errors of ±15° may be attributed to variations of the laser polarization axis, deviations in sample preparation, and the sample alignment in the SEM. We calculate NP area densities (NPs/µm$^2$) by analyzing the SEM images in MATLAB.

The presence of the self-aligned patterns only appears under the illumination of linearly-polarized laser light at the resonant frequency in regions of moderate NP area densities (6-16 NPs/µm$^2$). The chains of NPs are tilted but generally aligned perpendicular to the direction of the laser polarization. The perpendicular alignment to the linear polarization of the light is the signature of the Lorentz force dipole interactions, and these perpendicularly-aligned chains are independent of the orientation of the substrate. Hydrodynamic effects —such as thermophoresis and convection [30] [31], which have also been shown to produce self-assembly, albeit independent of the light polarization—are moderately suppressed in our system where the drying occurs on a hydrophilic surface and where laser light illuminates the entire drop (see Methods and Materials and Supplementary Materials for additional details).

In general, the structure of the line patterns depends on the area density of the NPs. In regions where the area density is 1-3 NP/µm$^2$, short, scattered lines are observed. These NP segments are less than 10 µm in length, but are interspersed uniformly. At area densities of 6-16 NP/µm$^2$ distinct patterns consisting of long parallel lines (> 100 µm) are observed, as shown in Fig. 3 (d-e). When the area density is higher, these longer NP strands appear more zigzagged. At the highest NP densities, heating, surface functionalization, and gradient-force attraction lead to the formation of blocks [32]. If the NP density is low (< 1 NP/µm$^2$), then the NPs are largely observed to be individually dispersed, and we infer that the larger separation between NPs results in Lorentz forces that are too small to overcome the random forces associated with Brownian motion.

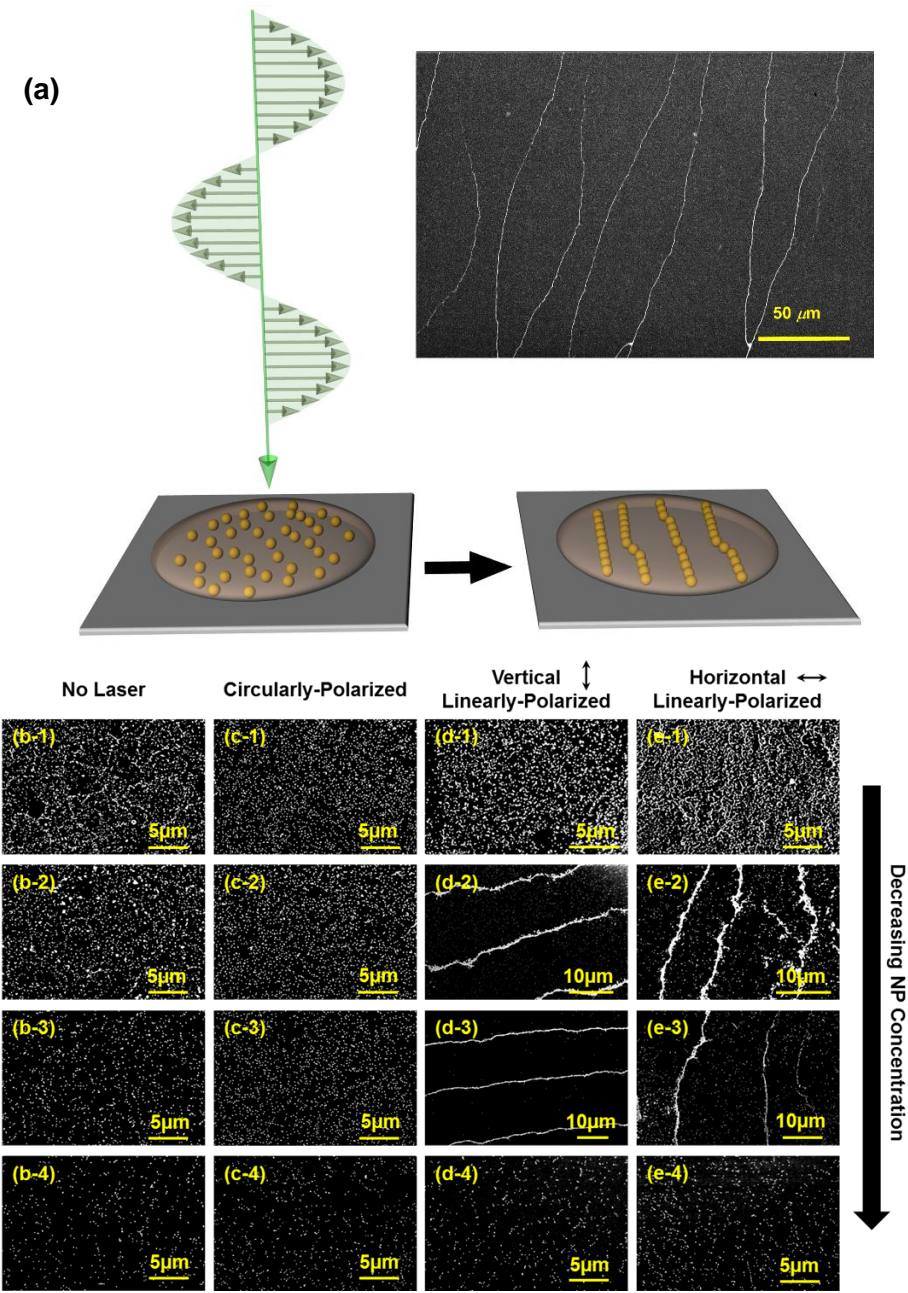

**Figure 3**. (a) Schematic diagram of NP drop under illumination of linearly polarized laser, and the SEM image of the string pattern of NPs. (b-e) SEM images of samples prepared (b) without laser (c) with circularly-polarized (d) vertical linearly-polarized and (e) horizontal linearly-polarized laser with NP concentrations (1) 0.0500 mg/mL, (2) 0.0375 mg/mL, (3) 0.0250 mg/mL, and (4) 0.0125 mg/mL. The laser has wavelength $\lambda$ = 532nm and peak laser intensity $I_{max}$ = 20MW/cm$^2$.

Strong solvent-substrate interactions are observed and are considered essential to the stability of the formed patterns [33]. Evidence of the substrate-solvent interactions is the presence of a distinct colloidal coffee ring, a diameter that identifies "pinning" of the initial droplet base [34] [See Fig. S1 (b)] that is marked by stacked Au NPs. We expect that the primary role of the substrate is to suppress the Brownian motion [35] [36]. In bulk NP solution, we measure the diffusion coefficient $D = 1.6 \cdot 10^{-12}$ m$^2$/s via dynamic light scattering, and the equivalent force is expected to be orders of magnitude larger than the mutual Lorentz force calculated here. In other words, we anticipate that in bulk solution, Brownian forces arrest the long-range pattern formation.

We also anticipate that electrostatic interactions between the substrate and sample have a strong role in promoting self-assembly; we observe that, where NPs aggregate into blocks or form thick lines, those NPs form a (single) mono-layer. The assembling process appears to commence after an initial group of NPs have been pinned to the substrate. To confirm this, we prepare samples with which we block the laser (linearly-polarized 532-nm) for a few minutes before the solution drop is completely dry. As expected, we observe no line patterns in the majority of these samples; those chains that do form manifest to a much lower degree.

We build a phase diagram that illustrates the pattern formation produced by varying illumination intensities and NP concentrations [Fig. 4 (a)]; we find that the phase plot varies depending on the imaged location of the drop (i.e., inside / outside the coffee ring). Experimentally, we observe both lower and upper NP-concentration thresholds for the self-aligned patterns, and we only observe a lower critical intensity threshold. With an intensity of 12 MW/cm$^2$, the ideal NP concentration is 0.0375 mg/mL, which corresponds to area densities of 11-16 NP/μm$^2$. The pattern formation only occurs at finite concentrations because the mutual Lorentz force competes with gradient and Brownian forces.

The degree of pattern formation increases with intensity; the NP self-alignment is observed over smaller areas in the droplet at lower intensity; as the intensity increases, the NP patterns are observed across larger areas of the drop. Fig. 4(b) shows the absorbance spectra from 0.0375mg/mL NP samples prepared with various illuminating laser intensities and polarizations. The absorbance spectra of the samples with the largest NP self-assembly show extra peaks at wavelengths of λ = 700 nm; the prominence of the 700-nm peak corresponds with the completion of NP self-assembly and is associated with the resonance of 1-D Au NP arrays or dimers [37]. From the relative intensity of the ~700 nm band and SEM sample scans, we estimate that at most 30% of the NPs are self-assembled into strings via Lorentz forces, and this ratio decreases with decreasing illumination intensity.

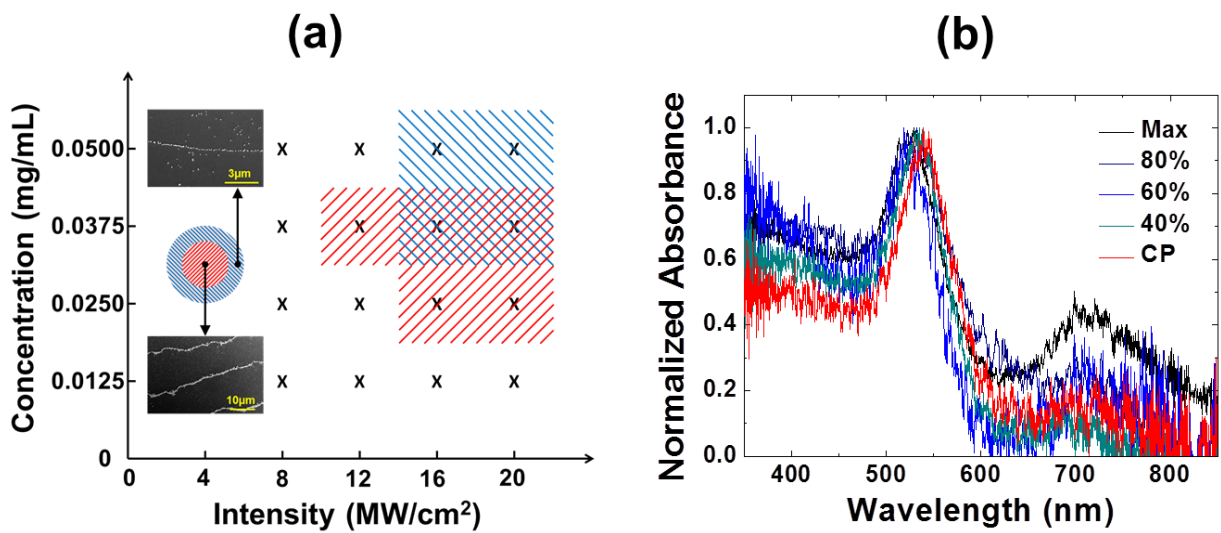

**Figure 4. (a)** Phase diagram of the Au NP self-assembly, which is observed as long parallel lines in the central region (red lines) and scattered short lines in outside region (blue lines), shown in the insets. **(b)** Normalized absorbance of samples prepared with 0.0375 mg/mL concentration (11-16 NP/$\mu m^2$): samples are illuminated by linearly-polarized laser light at a range of intensities (8 to 20 MW/cm$^2$) and circularly-polarized laser light at the maximum intensity (20 MW/cm$^2$).

We generalize that the long-range self-assembly mechanism of the NPs under resonant illumination occurs as follows. Initially NP chains are seeded by a dimer or a pair of attracted NPs. This pattern formation occurs in a manner analogous to spinodal decomposition and occurs uniformly throughout the sample, rather than at nucleation sites at the boundary of the sample. The NPs both settle at stable nodes produced by the mutual Lorentz force and also produce stable nodes via their radiation, migrating towards more stable positions that minimize the energy of the system. The NPs align perpendicular to the polarization and generally grow towards other self-assembled NPs that are also aligned on-axis. As NP patterns grow, they produce stronger dipole interaction forces, extending influence to other nearby NPs, which also align perpendicular to the laser polarization as they are drawn to the self-assembled NP-chain structures [as illustrated in Figs. 2 (c) and (d)]. This cascaded process of assembly, which leverages the intermediate configurations of stability, leads to NP chains that are orders of magnitude longer than the extent of the plasmon evanescent fields.

To summarize, we have investigated the non-negligible time-averaged mutual Lorentz force associated with plasmon radiation. These interactions have an accumulated long-range effect on the self-assembly of plasmonic NPs, particularly when dried on a flat substrate where the thermal Brownian forces are minimized. Experimental images of NP solutions dried on ITO substrate under linearly-polarized laser illumination at the plasmon resonance consistently reveal these effects. We observe a process of pattern formation that is analogous to spinodal decomposition, where disperse, aligned formations of individual dimers gradually join into longer NP strands as long as 200 μm, aligned perpendicular to the linear polarization of the illuminating laser. The longer-range responses are expected to emerge from the cascaded plasmon interactions and intermediate stable configurations dictated by the mutual Lorentz force, which is relevant to nascent models of plasmonically-powered processes.

**Acknowledgments.**

LTV graciously acknowledges funding from NSF DMR 1151783 and DMR 1709446. HJ is funded via the City University Advanced Science Research Postdoctoral Collaborative Grant. This work was performed in part at the Advanced Science Research Center NanoFabrication Facility of the Graduate Center at the City University of New York. The authors acknowledge helpful discussions with Benjamin W. Stewart and Michael Mirkin.